# The Meaning of Li diffusion in Cathode Materials for the Cycling of Li-ion Batteries: A Case Study on LiNi$_{0.33}$Mn$_{0.33}$Co$_{0.33}$O$_2$ Thin Films

Erwin Hüger[1,2,*], and Harald Schmidt[1,2]

[1]Clausthal University of Technology, Institute of Metallurgy, Solid State Kinetics Group, 38678 Clausthal-Zellerfeld, Germany.

[2]Clausthal Center for Materials Technology, 38678 Clausthal-Zellerfeld, Germany.

*Corresponding author: erwin.hueger@tu-clausthal.de

**Abstract**

We demonstrate that for polycrystalline LiNi$_{0.33}$Mn$_{0.33}$Co$_{0.33}$O$_2$ c-axis textured thin film cathodes of rechargeable lithium-ion batteries, the kinetics of Li storage and release including maximum specific capacity is determined by Li diffusion. The C-rate capability and long-term cycling behavior were investigated. The films exhibited up to 30% of the expected practical capacity even at low C-rates. However, 100% capacity was achieved at very low cycling rates below 0.01C. The capacity showed a reversible behaviour with changing current density, indicating no film degradation. The C-rate capability experiment showed a square root dependence of capacities on current density, which corresponds to a diffusion-controlled process. The estimated diffusivities from the cycling experiments are independent of the current density. The Li chemical and tracer diffusivities were measured using standard electrochemical and non-electrochemical diffusion measurement techniques. Chemical diffusivities, thermodynamic factor, and hence Li tracer diffusivities were determined from potentiostatic intermittent titration (PITT) and electrochemical impedance spectroscopy (EIS) experiments as a function of electrode potential and state of charge (SOC). The diffusivities were found to be approximately independent of potential, SOC and cycle number. The Li tracer diffusivities were validated by $^6$Li tracer diffusion experiments with secondary ion mass spectrometry (SIMS). The diffusivities obtained by PITT and SIMS were found to be more reliable for Li uptake and release than those obtained by EIS. Based on the diffusion results, a C-rate limit for full film delithiation below 0.01 C was calculated due to slow Li diffusion.

1. Introduction

Thin films are regarded as a model system for the investigation and improvement of active materials utilized in commercial Li ion batteries (LIBs) for the following six reasons.

First, thin films possess stable and flat interfaces that ensure a uniform electric field distribution and avert the formation of local hot spots. Such hot spots can boost Li dendrite formation and facilitate heterogeneous Li insertion and extraction across the electrode/electrolyte interface, complicating the interpretation of measurement data.

Secondly, thin films do not require conductive and binder additives, which are widely used in commercial LIBs in form of powders. A perspective on LIB cathode materials [1] asserts the necessity to study the electrochemical behavior of binder- and additive-free electrodes to comprehend the intrinsic properties of the material and to address the associated challenges [2]. Fabrication of thin film electrodes offers distinct advantages, as the electrochemical study of bulk materials is often hindered by electrode polarization and the necessity of conductive additives to mitigate large overpotentials. For example, in the development and optimization of electrode materials, it is necessary to quantify and understand the solid-state diffusion processes that govern charging and discharging [2]. The interaction of the electrolyte with the conductive network (binders and conductive additives), as well as the electrical contact with the conductive network throughout the electrode, complicates the determination of intrinsic diffusivities in LIB electrodes [2].

Thirdly, the study of thin film electrodes enables the implementation of sophisticated analytical methods of materials characterization. Materials exhibiting an exceedingly low electric conductivity, such as $LiNbO_3$ cannot be used in their bulk form as electrodes. Moreover, these materials are unsuitable for electron spectroscopy measurements (e.g. X-ray photoelectron spectroscopy) due to sample charging. It has been hypothesized that such "insulators" are inactive for Li insertion and extraction, a phenomenon that can be circumvented in the thin film configuration, as shown for LiNbO$_3$ thin films devoid of any conductive additives [3]. This phenomenon can be attributed to the charge transfer facilitation to the underlying conductive substrate by thin films, thereby minimizing surface charge accumulation, as described in reference [4]. However, the low number of electrons of Li severely restricts the measurement techniques capable of detecting and tracking Li in materials. Secondary ion mass spectrometry (SIMS) has been shown to have the capacity to detect lithium (Li) and differentiate between its stable isotopes, with natural abundances of 92.6% for $^7$Li and 7.4% for $^6$Li, respectively. The high ionization cross section for positive ions leads to the observation that SIMS secondary signals are of high intensity and detection sensitivity. However, it should be noted that SIMS does require relatively flat surfaces to properly measure the distribution of Li in the active material, as present in thin films.

Fourthly, thin film electrodes may be suitable for micro-batteries in wearable electronics, smart cards, wireless sensors, microelectromechanical systems, memory backups, e-textiles, implantable medical devices, and healthcare instruments [5-10]. This is not the case for the classical LIB design due to size, weight, and dimensional constraints. Thin-film batteries are inherently safe, scalable, easy integrable, and show a low self-discharge rate. Specifically, thin-film batteries exhibit the capacity to maintain full charging for extended periods spanning several years [8]. These attributes enable thin-film batteries to be well-suited for applications requiring long-term energy storage, such as medical implants and remote sensors.

Fifth, thin films show promise in overcoming the problems of large volume changes during cycling of active electrode materials. The next generation of LIBs should prioritize a high Li storage capacity, a task for which silicon is particularly well-suited for negative electrodes. The high Li capacity of silicon (theoretically up to Li$_{4.4}$Si, practically up to Li$_{3.75}$Si [11]) results in rapid loss due to pulverization and delamination. A comprehensive review of the literature (e.g. ref. [9]) enumerates additional advantages of thin films in LIBs, including their high surface area to volume ratio, which facilitates rapid lithiation and de-lithiation due to a shortened path for electron transport and Li diffusion.

Sixth, thin films offer the opportunity to adequately study cycling and Li kinetics in specific crystallographic directions. The mobility of Li within the binder and conducting phase surrounding the NMC powder particles is presumably high. Consequently, Li can enter the NMC crystalline particles via a preferred crystallographic direction, where Li insertion and extraction is more easily possible. A classic example are layered LiCoO$_2$ (LCO) and Li(Ni, Mn, Co)O$_2$ (NMC) cathode materials with an anisotropic crystal structure, supporting Li insertion perpendicular to the c-direction [12]. Further details are provided in section 1 and 2 of the supplemental material (SM) accompanying the present work. This phenomenon is not exclusive for layered cathode materials [12,13], but is also observed for crystalline silicon, where it is referred to as anisotropic lithiation and volume expansion [14]. The use of a slurry of microparticles (crystals) embedded in binders and additives complicates the exclusive measurement of Li insertion/extraction behavior as a function of the crystallographic direction. However, this phenomenon can be thoroughly investigated through the use of thin films or single crystals. While single crystals are rare, films with a thickness of a few microns often grow in columns perpendicular to the film surface. These films frequently exhibit textured properties, meaning they have a preferred orientation, such as c-axis oriented or ab-plane oriented NMC films. Li diffusion and cycling behavior can be adequately studied separately for both directions. Further details can be found in the second and third section of the SM to this paper.

The present study is an examination of Li diffusion in $LiNi_{0.33}Mn_{0.33}Co_{0.33}O_2$ (NMC111) binder-free and additive-free thin films. While the present study focuses on the C-rate and long-term cycling capability, its primary objective is not to investigate the rate performance. Instead, the study aims to elucidate the correlation between Li diffusivities and cycling behavior. The present study utilizes standard electrochemical methods, such as electrochemical impedance spectroscopy (EIS) and the potentiostatic intermittent titration technique (PITT) to investigate Li diffusion. A recent paper [15] concluded that the intrinsic Li diffusivity cannot consistently be identified by indirect diffusion determination methods based on electrochemical measurements. To solve this problem, it is necessary to determine the Li diffusivity independently using established diffusion determination methods for comparison, such as tracer diffusion with SIMS depth profiling experiments.

PITT and EIS were already applied to assess the Li diffusivity at ambient temperature in sintered LCO pellets without additives [15]. The determined Li diffusivities at room temperature range from $10^{-9}$ to $10^{-25}$ $m^2s^{-1}$ for the chemical Li diffusivity and from $10^{-9}$ to $10^{-28}$ $m^2s^{-1}$ for the tracer Li diffusivity [15]. The results of the SIMS diffusion experiments help to limit the range of possible electrochemically determined Li tracer diffusivities to a narrow band within an order of magnitude around $10^{-22}$ $m^2s^{-1}$ [15]. Consequently, in the current work, the standard SIMS diffusion measurement techniques were also applied with Li isotope depth profiling for comparison.

A review on the investigation of Li diffusivities in NMC, with a focus on additive-free NMC111 thin films is presented in section 3 of the SM. A comprehensive study was conducted by Strafela et al. [16] on the effect of different magnetron sputtering parameters on the NMC111 film orientation and cycling behavior such as argon partial pressure. The study involved the deposition of 1 µm thin NMC111 films on stainless steel substrates, a method comparable to that employed in the present study. X-ray diffraction (XRD) and transition electron microscopy (TEM) analysis revealed the presence of columnar grains with a preferential c-axis orientation (hexagonal (0001) planes parallel to the surface) [16]. Notably, Strafela et al. reported a remarkably low gravimetric capacity for the NMC111 films down to 0.047 mAhg$^{-1}$ [16], representing only 0.017% of the theoretical capacity of NMC111 (approximately 273 mAhg$^{-1}$) [18]. The cycling behavior of LCO thin films prepared by pulsed laser deposition (PLD) with a preferred c-axis orientation was compared to non-c-axis oriented films, indicating different electrochemical properties [17]. The c-axis oriented LCO films of ref. [17] are similar to the NMC111 films of this study in the following way: Firstly, the XRD patterns (described in section 4 of the SM) are similar [17], exhibiting an increased (003) reflection relative to the other reflections. Further, the c-axis oriented films exhibited an unexpected property, namely diminished reactivity to Li during cycling, resulting in lower capacities when compared to non-c-axis oriented films. The underlying reasons for this behavior remain unanswered in the literature and require further investigation as mentioned in reference [17]. The present experimental work aims to contribute to this ongoing investigation.

The present study encompasses three distinct segments of research. The first segment pertains to the preparation of ion-beam sputtered crystalline NMC films as LIB electrodes (section 2), the characterization of their crystal and defect structure (section 4 in the SM), and their cycling behavior (section 3.1). It was observed that these electrodes exhibit a peculiar cycling behavior with reduced specific capacity. The subsequent two blocks of work are based on Li diffusivity determination, which was performed to understand the cycling behavior. In the second block of work, electrochemical measurement techniques were applied to determine the chemical diffusivity during cycling (section 3.2) by performing EIS experiments (section 3.2.1) and PITT experiments (section 3.2.2). The PITT technique was also employed to ascertain the thermodynamic factor that links the chemical to the Li tracer diffusivity. To validate the diffusivities obtained by EIS and PITT, SIMS experiments were conducted to determine the Li tracer diffusivity and the activation enthalpy of Li diffusion in NMC film electrodes (section 3.3). This constitutes the third block of work. Section 4 discusses the obtained Li chemical diffusivities in section (4.1), the Li tracer diffusivities in section (4.2), and the influence of the Li diffusivity on the electrode specific capacity in section (4.3). The results are summarized in section 5.

The supplementary material to this paper provides further details on the following subjects: (1) Orientations and crystallographic directions in hexagonal layered phases, (2) cycling performance and Li diffusivity at certain crystallographic directions, (3) literature review on cycling behavior and determination of Li diffusivity in additive-free NMC111 thin films, (4) characterization of NMC111 film electrodes; (5) SIMS standard technique of Li diffusivity determination applied to NMC and LCO, (6) diffusion-controlled cycling studied by C-rate capability experiments, (7) PITT experiments and further electrochemical results of cycling the 1 µm thin NMC111 film electrode. (8) EIS results on the 1 µm thin NMC111 film electrode, (9) electrochemical results of a 500 nm NMC111 film electrode, and (10) additional references.

## 2. Materials and Methods

Thin films of NMC111 were deposited by ion beam sputtering on polished cylindrical stainless steel (SS) substrates with a diameter of 1 inch and a thickness of 1 mm. The SS substrate served as the current collector and the NMC film as the active material of the working electrode (WE), hereafter referred to as the NMC111 film electrode.

Ion-beam sputtering was performed using a commercial setup (IBC 681, Gatan, Irvine, CA, USA) equipped with two Penning ion sources. Deposition was performed at 5 keV and 200 µA using argon sputtering gas at an operating pressure of $5 \times 10^{-3}$ mbar. The NMC111 sputtering targets were prepared by solid-state syntheses using a two-step reaction route. $^{nat}Li_2CO_3$ or $^6Li_2CO_3$ (95% $^6Li$ enriched) was obtained from Sigma-Aldrich (Taufkirchen, Germany), and NiO, $Co_3O_4$, and $MnO_2$ powders were obtained from Carl Roth (Karlsruhe, Germany). The starting materials were mixed in the appropriate stoichiometric ratio in an agate ball mill (Wisd WiseMix Ball Mill, Wisd Laboratory Instruments, Wertheim, Germany) by adding an appropriate amount of dispersing agent (ethanol) at 150 rpm for 1 h. The black slurry is completely dried and the resulting powder mixture is heated to 800 °C at 3 K/min in pure oxygen for 16 h. After subsequent high-energy ball milling of the powder mixture (SPEX 8000M shaker mill, SPEX, CertiPrep, SampleSpec, Metuchen, NJ, USA) for a total of 4 min at 1080 cycles/min with zirconia ceramic balls, a 20 mm diameter pellet was pressed at 330 MPa and sintered at 800 °C in air for 12 h, yielding a polycrystalline dense sputtering target. To determine the relative element concentrations, the sputter targets were analyzed by inductively coupled plasma optical emission spectroscopy (ICP-OES) using a Plasma Quant 9100 (Analytik Jena, Jena, Germany). A composition of $Li_{0.97}(Ni_{0.31}Mn_{0.32}Co_{0.34})O_{2.05}$ was obtained, giving a Li to metal ratio close to 1. Since it is expected that there is a loss of a few single percent of Li during the sputter deposition process, we expect a slight sub-stoichiometry of the thin film samples. The as-deposited films were annealed at 700 for 1h in air in order to form a crystalline state.

Electrochemical measurements were performed using a three-electrode electrochemical cell. Propylene carbonate (PC, Sigma Aldrich, Taufkirchen, Germany, anhydrous, 99.7%) with 1 M $LiClO_4$ (Sigma Aldrich, Taufkirchen, Germany, battery grade) was used as the electrolyte. No separator was used due to the large distance of 20 mm between the electrodes. The cell was assembled and disassembled in a glove box filled with argon gas, and the partial pressures of $O_2$ and $H_2O$ were less than 1 ppm. The diameter of the lithiated zone was determined visually on the electrode surface to about 15 mm after disassembling the cell, and coincides with the diameter of the cell tube filled with liquid electrolyte. The counter and reference electrodes were lithium plates (1.5 mm thick, 99.9%, Alfa Aesar, Kandel, Germany). All reported NMC potentials (Ewe) refer to the Li metal reference electrode. The process of $Li^+$ extraction (delithiation) and insertion (lithiation) from the NMC electrode corresponds to cell charging and discharging, respectively. In this article, the terms delithiation and lithiation will be used instead, although the terms charging and discharging are ubiquitous in the literature. Each cycle begins with delithiation of the NMC electrode. Electrochemical studies were performed on a Biologic SP150 potentiostat and the EC-lab software version V11.43 (Biologic, Seyssinet-Pariset, France).

In this work, we focus on the determination of diffusivities. Diffusion in solids is often a slow process at room temperature. Therefore, PITT and EIS measurements were carried out in a way to adequately determine the diffusivity. The PITT measurements were performed down to low $Li^+$ current densities (1 $\mu Acm^{-2}$) over long time intervals. For EIS, the mass transfer impedance response is expected to appear in the very low frequency range, and the acquisition of EIS spectra is time-consuming up to days for measurements well below 1 mHz [19]. Unfortunately, the long acquisition time increases the influence of unwanted external influences on the EIS data [20]. We decided to measure EIS down to a frequency of 1 mHz with an acquisition time of 7 hours per spectrum. Potentiostatic EIS was performed with an amplitude of 10 mV. All electrochemical and SIMS measurements were performed at room temperature.

For the Li tracer diffusion experiments, tracer deposition was performed by depositing a thin layer of $^6Li$ isotope-enriched NMC111, 50-90 nm thick, on pristine or cycled NMC111 film electrodes of about 1.5 µm thickness by ion-beam sputtering. Since the $^6Li$ tracer NMC111 layer and the $^{nat}Li$ NMC111 film have approximately the same chemical composition (same production route), pure isotope interdiffusion is expected during the diffusion experiments and then tracer diffusivity can be determined.

Depth profiles of the two Li isotopes and of Ni, Mn and Co were obtained by SIMS using a Cameca ims 3f/4f machine (CAMECA SAS, Gennevilliers Cedex, France) with (15 keV, 100 nA) O$^-$ primary ions. Analysis of the resulting sputter craters was performed using a mechanical stylus profilometer (Tencor Alphastep 500, Milpitas, CA, USA). The relative $^6$Li isotope fraction, $c$, is calculated from the $^6$Li and $^7$Li signal intensities, $I$, measured by SIMS, as follows

$$c = \frac{I(^6Li)}{I(^6Li)+I(^7Li)}. \qquad (1)$$

The structural investigation of the NMC film samples was performed by grazing incidence X-ray diffraction (GI-XRD) measurements using a Bruker D8 Discover diffractometer (Bruker AXS Advanced X-ray Solutions GmbH, Karlsruhe, Germany) with CuK$_α$, 40 keV, 40 mA, angle of incidence 2°. GI-XRD investigation of the ion-beam sputtered 1 µm thin NMC111 films annealed at 700°C for one hour revealed a layered NMC phase with a rather low amount of Li/Ni cation mixing. It is further shown that the NMC111 films are c-axis textured, where the grains have the hexagonal Li layers parallel to the film surface. Therefore, Li extraction and insertion should be perpendicular to the hexagonal lithium, oxygen and transition metal layers, as shown in Figure S2b in the SM. Further details are given in Sections 1, 2, 3, and 4 of the SM to this paper and in the Discussion in Section 4 of this paper.

## 3. Results

### 3.1 Cycling Performance

Electrochemical investigations were done predominantly on a crystalline NMC111 film electrode with a thickness of 1 µm. PITT experiments were carried out during the first cycle. The charge was extracted during the PITT experiment by constant voltage steps over a long overall time-interval of 113 hours, which corresponds to a very low C-rate of 0.008C. The total extracted charge density is 91 µAhcm$^{-2}$ and the gravimetric capacity is 227 mAhg$^{-1}$. This capacity is lower than the theoretical capacity of NMC111 (≈ 273 mAhg$^{-1}$), but higher than the practical capacity (≈ 165 mAhg$^{-1}$) [18]. The re-lithiation process of the film electrode, which is a discharge process in the battery, already took place during the open circuit period due to self-discharge as described in Section 7.1 of the SM. Therefore, PITT could not be performed during the lithiation process of the first cycle.

The second cycle was devoted to the measurement of numerous EIS spectra, each at a given constant potential during the delithiation process and the subsequent lithiation process. The charge capacity during delithiation decreases by ≈ 50% from the first to the second cycle, from 227 mAhg$^{-1}$ to 97 mAhg$^{-1}$. The capacity of 0.04 mAhg$^{-1}$ for the re-lithiation process of the second cycle remains unexpectedly low. More details are given in sections 7 and 9 of the SM.

After the first two cycles, the electrode performance was tested with a constant current (CC) C-rate capability experiment in the potential window between 1.7 V and 4.6 V (Figure 1) for current densities between 0.4 µAcm$^{-2}$ and 40 µAcm$^{-2}$. The measurement data are shown in Figure S6 in the SM of this paper. A cycling protocol was used to test rate capability by incrementally increasing the current density from 0.4 to 40 µAcm$^{-2}$ after each fifth cycle and then decreasing the current density back to 0.4 µAcm$^{-2}$ (Figure 1). Afterwards, 278 additional CC cycles at medium constant current density of 4.0 µAcm$^{-2}$ was done (Figure 2).

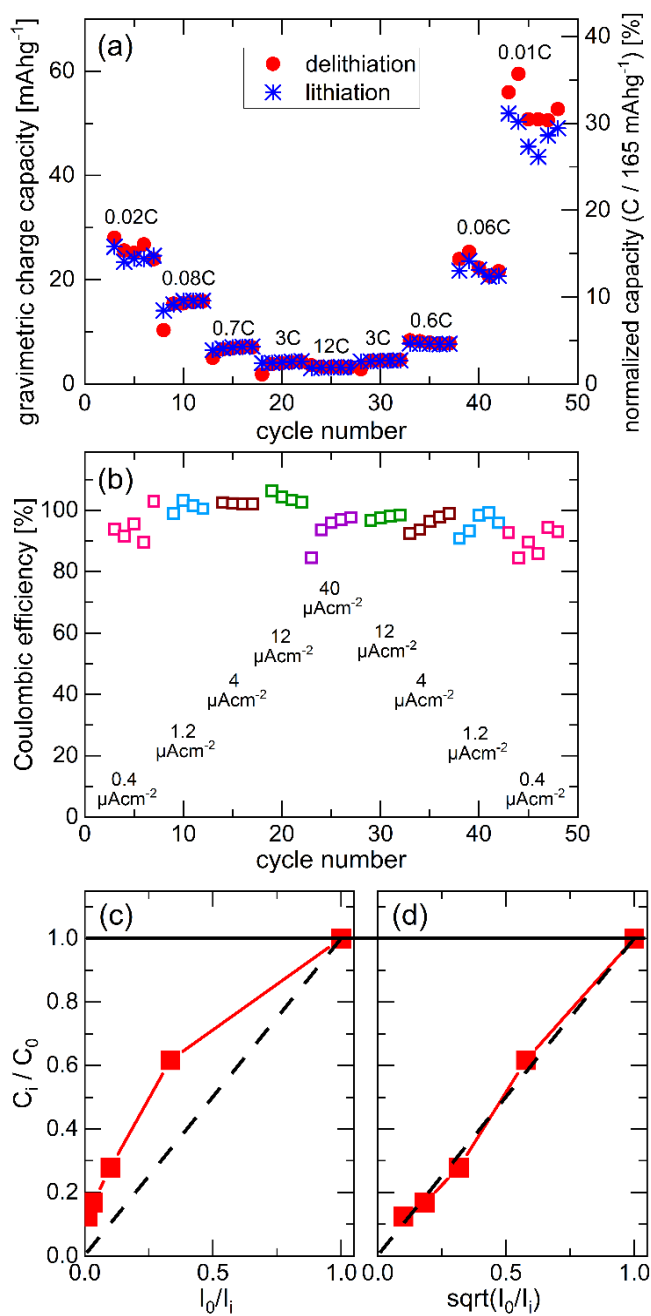

**Figure 1**. C-rate capability of the 1 µm NMC111 film. (a) Gravimetric charge and discharge capacity and (b) Coulombic efficiency for constant-current Li$^+$ insertion and extraction cycles performed at the current densities shown in Figure 1b (the cycling protocol). The delithiation (charging) and lithiation (discharging) capacities are represented as red dots and blue snowflakes, respectively, in panel (a). The color in panel (b) is only for better visual differentiation. The C-rate shown in (a) was calculated by dividing 1 h by the lithiation time interval given in hours. (c,d) Plots show the normalized decrease in capacity with increasing normalized Li current density. Details are given in the main text.

Figure 1a shows a successive decrease in capacity with increasing current density. A possible reason might be that the NMC111 film degrades, meaning some parts of the film separate from the current collector and do not participate in further cycling. However, this is not the case here because subsequent cycling at lower currents (Figure 1a,b) not only restore the capacity in a stepwise and reversible manner, but even doubles the capacity to ≈ 50 mAhg$^{-1}$ at cycle 43 to 48 compared to ≈ 25 mAhg$^{-1}$ at cycle 3 to 7. This is also observed in the long-term cycling experiments done afterwards at a current density of 4 µAhcm$^{-2}$ (0.7 C) for additional 277 cycles (Figure 2). During cycling, the capacity increases from ≈ 4 mAhg$^{-1}$ in cycle 18 to 22, to ≈ 7 mAhg$^{-1}$ in cycle 33 to 37, and finally to ≈ 13 mAhg$^{-1}$ in cycle 300. When cycling back to the lowest current density, the capacity returns to ≈ 53 mAhg$^{-1}$ at cycles 326 and 327. Thus, the C-rate capability (Figure 1) and long-term cycling (Figure 2) experiments show no evidence of film degradation. We suggest that the decrease in capacity at higher current densities is most likely due to limitations caused by slow Li diffusion.

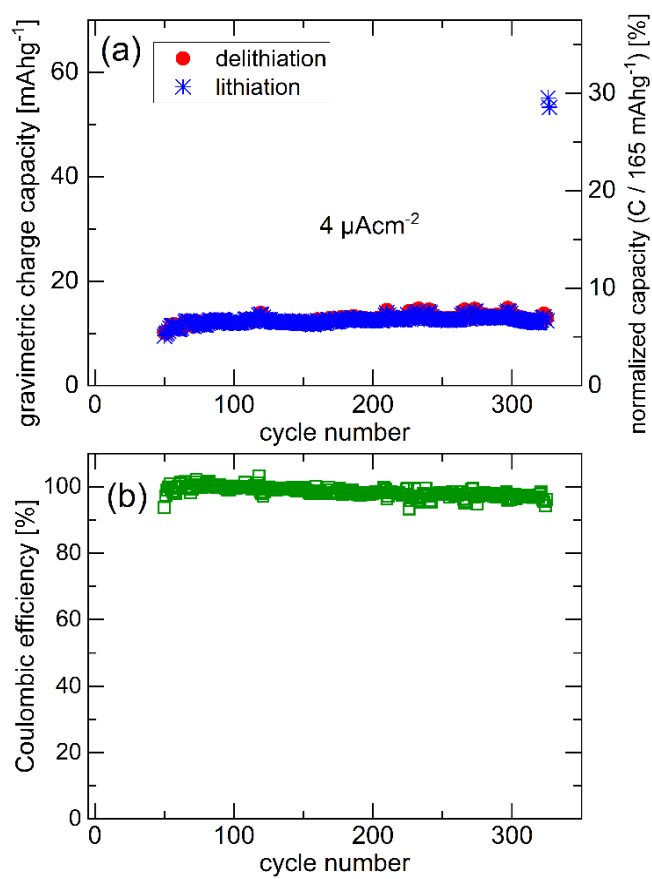

**Figure 2.** Long-term cycling study. Gravimetric capacity (a) and Coulombic efficiency (b) of the 1 µm thin NMC111 film as a function of the number of cycles. The delithiation (charging) and lithiation (discharging) capacities are represented by red dots and blue snowflakes, respectively, in panel (a).

Figure 1c,d examines whether Li diffusion can be a determining factor for the C-rate capability. For simplicity, only the forward branch of the cycles (the current increase) is considered up to circle number 28 in Figure 1a during delithiation. The red squares in Figure 1c,d show the fraction $C_i/C_0$ between the mean discharge capacity $C_i$ (averaged over five cycles) at each current density $I_i$ normalized to the mean capacity $C_0$ obtained at the lowest current density of 0.4 µAcm$^{-2}$ at $I_0$. The deviation of the measured capacity $C_i$ at each current density $I_i$ from the ideal case ($C_i = C_0$) is illustrated in Figure 1c,d. The ideal case means that the capacity does not depend on the current density and is indicated with a solid horizontal line in Figure 1c,d at $C_i/C_0 = 1$. For the ideal case, at higher current densities the delithiation time has to be shorter, according to $Q = I_i \cdot t_i = I_0 \cdot t_0$. Figure 1d shows that the capacity decrease at higher currents depends on the square root of the inverse current density. In section 6 of the SM of this paper, the following equation is developed for the case where the capacity is limited by Li diffusion

$$\frac{C_{i+1}}{C_i} = \frac{\Delta Q_{i+1}}{\Delta Q_i} = \sqrt{\frac{I_i}{I_{i+1}}} \qquad (2)$$

Here, $\Delta Q_i$ is the Li charge stored or removed during the CC cycle with the current density $I_i$.

Equation (2) is similar to the well-known case of cyclovoltammetry (CV) experiments performed with different sweep rates $v_i$ [3,21], where in the case of diffusion-controlled Li insertion and extraction the following equation determines the CV experiment.

$$\frac{C_{i+1}}{C_i} = \frac{\Delta Q_{i+1}}{\Delta Q_i} = \sqrt{\frac{v_i}{v_{i+1}}}. \qquad (3)$$

Equations (2,3) define a purely kinetic regime. For instance, there is an analogy between a diffusion-controlled behavior of capacity, charge and current obtained from CV experiments with different sweep rates (equation (3)) and a diffusion-controlled behavior of capacity and charge obtained from C-rate experiments with different current densities (equation (2)). Thus, the slow Li diffusion in the NMC111 film is probably the reason for the capacity decrease during the C-rate test according to equation (2) and Figure 1c,d. The cycling behavior, i.e., the specific capacity, is determined by Li diffusion. That implies that a reduced capacity at higher current densities might be the consequence of an incompletely filled electrode due to slow Li diffusion.

To further test a possible dependence of the NMC111 film capacity on Li diffusivities, the results obtained on the 1 µm film are compared to those obtained on a 500 nm NMC111 film. This is described and discussed in Section 9 of the SM (Figure S20c). If the Li diffusivity is high enough to delithiate the NMC111 film completely, and the specific capacity is independent of the film thickness the charge extracted from the 500 nm thin film should be smaller than that from the 1 µm NMC111 film. The comparison is made in section 9 of the SM at a current density of 12 µAcm$^{-2}$ (Figure S20c). However, the charge extracted from the thinner film is not lower than that from the thicker film. This indicates that delithiation does not take place within the whole film electrode, but only in a limited part of the electrode close to the electrolyte, i.e. from a depth well below the film thicknesses of 1000 and 500 nm.

For the 1 µm thick NMC111 film, the amount of charge density of 2 µAhcm$^{-2}$ extracted during delithiation (Figure S20c) is only 3% of the maximum charge density that can be extracted (66 µAhcm$^{-2}$) considering the reported practical capacity for NMC ($\approx$165 mAhg$^{-1}$) [18]. It is likely that only a small part of the NMC111 electrode is delithiated/lithiated due to a presumably low diffusivity. From this point of view, a Li diffusion length of 3% of the NMC111 film thickness of about d = 30 nm is roughly estimated. CC delithiation at the operating current density of 12

µAcm$^{-2}$ took t ≈ 0.36 hours. Therefore, using D=d$^2$/(2t), the Li diffusivity is estimated to be ≈ 3 × 10$^{-17}$ m$^2$s$^{-1}$ in good agreement to the results of the PITT experiments discussed below.

All the cycling tests performed (Figure 1, Figure 2, and Figure S20) indicate that low Li diffusivities may be the reason for the poor cycling performance. Therefore, it is appropriate to study the Li diffusion in the NMC111 film at all SOCs.

During the last two cycles 328 and 329 (before the electrochemical cell was disassembled for the SIMS experiments), PITT and EIS experiments were performed again to measure the diffusivity electrochemically after long-term cycling. A long-term PITT experiment was performed in cycle 328 (details are given in section 3.2.2 of this paper and in section 7 of the SM). The total charge density extracted during the PITT experiment in cycle 328 was 125 µAhcm$^{-2}$, which corresponds to a gravimetric capacity of 312 mAhg$^{-1}$, similar to the first cycle PITT experiment and again higher than the theoretical maximum capacity of NMC (273 mAhg$^{-1}$). The charge density introduced during the PITT experiment in the re-lithiation process of cycle 328 was 40 µAhcm$^{-2}$, corresponding to a gravimetric capacity of 100 mAhg$^{-1}$, which is lower than the practical capacity of NMC (165 mAhg$^{-1}$). Finally, in the last cycle (cycle 329), EIS experiments were performed during the potential hold at the NMC111 film electrode. During delithiation, the total extracted charge density was 14 µAhcm$^{-2}$, corresponding to a gravimetric capacity of only 35 mAhg$^{-1}$. During re-lithiation, the total charge density introduced was 7 µAhcm$^{-2}$, corresponding to a gravimetric capacity of only 17.5 mAhg$^{-1}$.

## 3.2 Li Chemical Diffusivity Determination with Electrochemical Methods

### 3.2.1 Diffusivity Determination with EIS

EIS investigations were performed on the pristine 500 nm and 1 µm thin NMC111 film electrodes before any Li extraction, during cycle 2, and after the long-term cycling during cycle 329. Examples of typical results and details of the measurements, analysis and how the diffusivities were obtained from the EIS measurements are given in Sections 8 and 9 of the SM accompanying the current work.

Figure 3 shows the chemical diffusivities obtained from the EIS measurements. The diffusivities obtained during cycle 329 are plotted with red stars, together with those obtained during the second cycle marked with green squares and the value obtained from the pristine electrode, which is marked with a black triangle. Also shown are diffusivities obtained on the 500 nm sample. Most of the diffusivities are in the range of ≈ 1×10$^{-22}$ m$^2$s$^{-1}$ to ≈ 1×10$^{-25}$ m$^2$s$^{-1}$ not systematically changing with electrode potential up to 4.5 V, and cycle number. They are low diffusivities, consistent with the indications from the cycling experiments that low Li diffusivities determine the capacity. Further discussion is given in Sections 8 and 9 of the SM and in the Discussion section below.

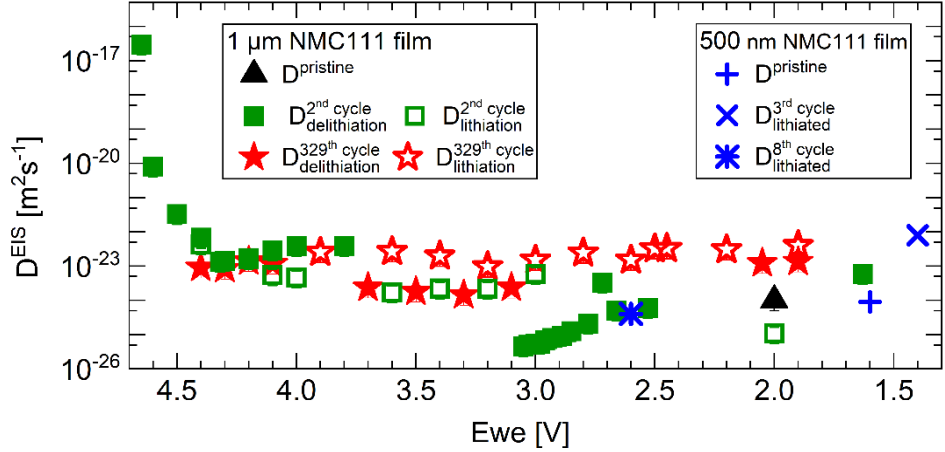

**Figure 3.** Diffusivities obtained from EIS experiments on the 1 µm thin NMC111 film electrode in the pristine (non-cycled) state (black triangle), during the second cycle (green squares) and during the 329[th] cycle (red stars) as a function of the electrode potential. The blue cross, the blue time sign, and the blue snowflake mark the diffusivities obtained from EIS measurements on the 500 nm thin NMC111 film electrode in the pristine state, in the re-lithiated state after the third cyclic voltammogram and in a 50% SOC state after another 8 cycles, respectively.

*3.2.2 Diffusivity Determination with PITT*

PITT experiments were performed during the first and 328[th] cycle. The chemical diffusivity and the thermodynamic factor were determined from the PITT experiments as described and critically evaluated in reference [22] and presented in Section 7 of the SM accompanying this work.

The charge density extracted during each PITT experiment is plotted in Figure 4a and Figure 4c, respectively. During the first cycle, charge density extraction starts around an NMC111 potential of 3.6 V, similar to the case of sintered NMC111 pellets [22]. The extracted charge density increases significantly above 3.6 V, reaching maxima at potentials of 3.9 V, 4.45 V, and 4.62 V. A broad peak between 3.7 and 4.0 V has also been reported for cyclovoltammetry measurements on NMC electrodes, including NMC111 [13,23-25]. The peaks occurring at the higher voltages are attributed to different crystallographic phase transitions associated with mechanical and electrochemical degradation (instability) [26], e.g. with exothermic release of oxygen and significant pressure build-up in the cell [25,26]. However, during delithiation of the 328[th] cycle, the charge density released at high potentials (around 4.5 V) is still present and even increases, indicating a reversibility of this delithiation process from the first to the 328[th] cycle. More details are given in Section 7 of the SM.

The corresponding diffusivities are shown in Figure 4b and 4d. The diffusivities are higher by orders of magnitude than those obtained from EIS. In contrast to the EIS results (Figure 3), a continuous decrease with increasing potential is seen for the PITT results (Figure 4b,d), especially between 3.5 and 4.0 V. Further details are given in Section 7 of the SM and in the Discussion section (Section 4) of this paper.

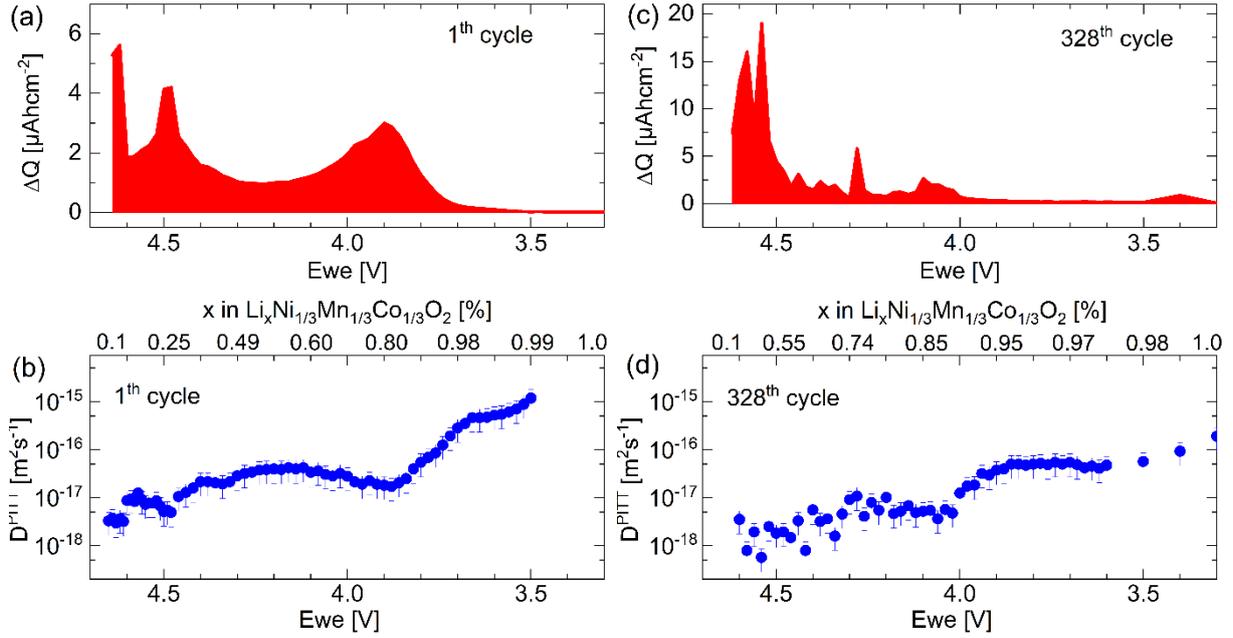

**Figure 4.** Results of the PITT experiment performed during the first **(a,b)** and the 328[th] **(c,d)** delithiation cycle of the 1 μm thin NMC111 film electrode as a function of the electrode potential and charge density state, expressed as x in $Li_xNi_{1/3}Mn_{1/3}Co_{1/3}O_2$. (a,c) The extracted charge density. (b,d) Diffusivities.

## 3.3 Li Tracer Diffusivity Determination with SIMS

SIMS experiments were performed to determine the Li tracer diffusivity in NMC111 film electrodes before and after all electrochemical experiments (after the 329[th] cycle). The measurements were done at room temperature for comparison with the electrochemically determined diffusivities and also at higher temperatures to determine the activation enthalpy of diffusion. A classical SIMS experiment on NMC films is described in [27]. The main point is that the re-distribution of the $^6$Li tracer after storing the sample at a certain temperature for a defined time is measured by SIMS. From the modification, the Li diffusivity can be determined (see below).

For the pristine NMC111 film electrode, it was not possible to extract a reliable Li diffusivity at room temperature by SIMS due to a very low diffusivity that did not result in sufficient modification of the tracer profile. However, assuming that the diffusion mechanism does not change with temperature, there is a way to obtain the diffusivity by extrapolation of diffusivities determined at higher temperatures. Diffusion is a thermally activated process with higher diffusivities at higher temperatures, where successful data were extracted.

The diffusivities of the crystalline films in the pristine state at elevated temperatures up to 200 °C are shown in Figure 5 with unfilled blue triangles [27]. They follow an Arrhenius law with an activation enthalpy of diffusion of (0.95 ± 0.08) eV and a pre-exponential factor of $\ln(D_0/m^2s^{-1}) = (-15.4 \pm 2.4)$.

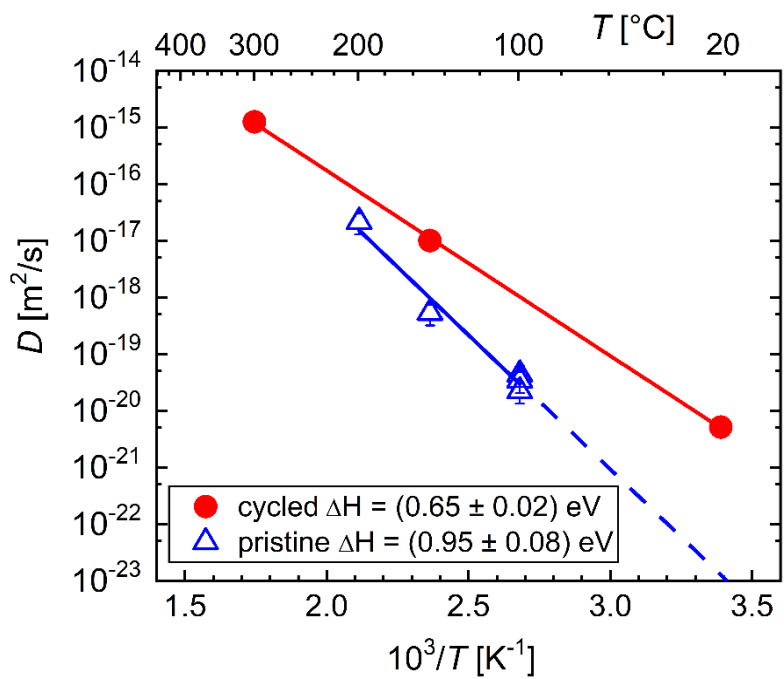

**Figure 5:** Temperature dependence of Li tracer diffusivities in pristine (i.e., electrochemically untreated) NMC111 films (unfilled blue triangles) and cycled NMC111 films (i.e., after the 329th cycle) (red dots). The solid lines represent fits using the Arrhenius law. The dashed line is an extrapolation to room temperature. The activation enthalpy of diffusion ΔH is indicated.

The range of Li tracer diffusivities in NMC111 extrapolated to room temperature (dashed line in Figure 5) is between $4 \times 10^{-23}$ m²s⁻¹ and $1 \times 10^{-23}$ m²s⁻¹ taking into account the error limits of the activation enthalpy and the pre-exponential factor. From our experience with SIMS Li depth profiling experiments on NMC materials, we conclude that diffusion lengths of at least d ≈ 50 nm can be well resolved. Considering the above estimated Li diffusivity of $D = 1 \times 10^{-23}$ m²s⁻¹, an estimated diffusion time interval of about t = d²/(2D) ≈ 4 years is obtained. This explains why a reliable diffusivity determination at room temperature for the pristine NMC111 film electrodes was not possible.

After the lithiation process of the 329th cycle, the open circuit voltage reached a value of 2.5 V, (see Figure S19), which corresponds to the last electrochemical measurement before the electrode was removed from the electrochemical cell and stored in an argon gas environment in a glove box. The electrode was rinsed with isopropanol for approximately 15 minutes, dried, and a 80 nm thin ⁶Li enriched NMC111 layer was sputter deposited on top. The electrode was then measured by SIMS and afterwards stored at room temperature for 149 days. The corresponding depth profiles can be seen in Figure 6.

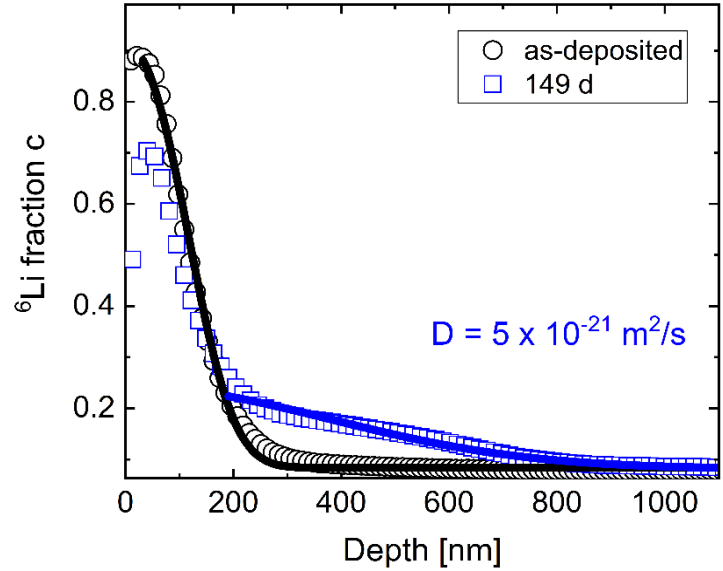

**Figure 6.** $^6$Li fraction as a function of depth for an electrochemically cycled sample (after 329$^{th}$ cycle) after tracer deposition and the same sample stored at room temperature for 149 days. Also shown are fitting curves according to equation (4).

Figure 6 shows the relative $^6$Li fraction as a function of depth before and after storage at room temperature. A broadening of the profile due to diffusion is visible. Immediately after tracer deposition, the depth profile can be fitted to Fick's second law for diffusion across an interface using the thick film solution [27]

$$c(x,t) = c_\infty + \frac{(c_0-c_\infty)}{2}\left[erf\left(\frac{h+x}{R}\right) + erf\left(\frac{h-x}{R}\right)\right], \tag{4}$$

where $c_\infty \approx 0.08$ is the residual $^6$Li concentration in the bulk NMC111 sample, $c_0 = 0.94$ is the $^6$Li concentration in the top $^6$Li enriched NMC111 layer, $h \approx 90$ nm is the thickness of the $^6$Li enriched layer, and $R_0 = 94$ nm is the broadening of the Li distribution at the interface. After annealing we obtain $R = 524$ nm and the diffusivity can be calculated as $D = (R^2 - R_0^2)/4t$, where $t = 149$ days. We obtain a higher diffusivity of $D = (5 \pm 2) \times 10^{-21}$ m$^2$s$^{-1}$ which is compared to the PITT, EIS, and literature values in the next section.

Corresponding SIMS experiments were also performed et elevated temperatures to determine the activation enthalpy of Li diffusion in the cycled NMC111 film electrodes. The diffusivities are shown in Figure 5 with red dots. The diffusivities of the cycled film are higher than those of the pristine film below 200 °C. This is mainly due to the lower activation enthalpy of diffusion of $(0.65 \pm 0.02)$ eV. The different activation enthalpies of diffusion indicate that the Li tracer diffusion mechanism changed from the pristine to the cycled NMC111 film electrode. In the cycled electrode, Li diffusion takes place along migration pathways with lower energy barriers.

It was found that the diffusivities in crystalline thin films (pristine electrodes of the current work) and in sintered bulk materials were found to be almost identical [27]. Furthermore, a comparative study of Li tracer diffusion in single crystals and polycrystalline LCO [12] showed that the ab-plane Li diffusivities of single crystals are the same as those observed in polycrystalline LCO. This finding suggests that grain boundaries have little influence on Li ion migration within the material. Along the c-axis, the diffusivities of single crystals are significantly lower. This provides experimental evidence for the often postulated slow Li diffusion along the c-axis.

## 4 Discussion on Diffusivities and their Influence on Specific Capacity

Figures 7 to 13 compare and summarize all results obtained from the EIS, PITT and SIMS studies on the 1 µm thin NMC111 film in comparison to literature values. The diffusivities obtained from EIS and PITT are obtained during the whole range of SOCs, whereas the SIMS investigations were performed only before the first cycle and after the lithiation process of the 329$^{th}$ cycle. First, the chemical diffusivities obtained from the PITT and EIS experiments are discussed in Section 4.1. Section 4.2 discusses the tracer diffusivities as obtained from SIMS experiments and from PITT and EIS measurements using the thermodynamic factor obtained from PITT experiments. Section 4.3 compares the diffusivities from the C-rate capability experiment and from the long-term cycling experiments with the diffusivities obtained from PITT, EIS, and SIMS.

### 4.1 Discussion of Chemical Diffusivities

The NMC111 film electrode potential (Ewe) as a function of SOC is shown in Figure 7a for the first cycle and in Figure 8a for the 328$^{th}$ cycle. The SOC was calculated from the amount of charge extracted and inserted during the PITT experiments as described by equation (s11) of the SM. It is observed that the charge density is extracted from the NMC111 film (i.e. the SOC decreases) only for potentials above 3.6 V. Lithiation takes place at much lower potentials than the delithiation process and shows a strong hysteresis (see Figure 8a), which is typical for active electrode materials with hindered Li$^+$ insertion extraction due to low Li diffusivities.

Figures 7b and 8b show the diffusivities obtained from PITT for the first and 328$^{th}$ cycle and from EIS for the second and 329$^{th}$ cycle as a function of SOC in comparison. The D$^{EIS}$ value obtained from the pristine NMC film electrode (i.e., corresponding to SOC = 100%) reaches ≈1×10$^{-24}$ m$^2$s$^{-1}$ (Figure S18b in the SM) and is overlapping with the D$^{EIS}$ values obtained at the beginning of the second cycle.

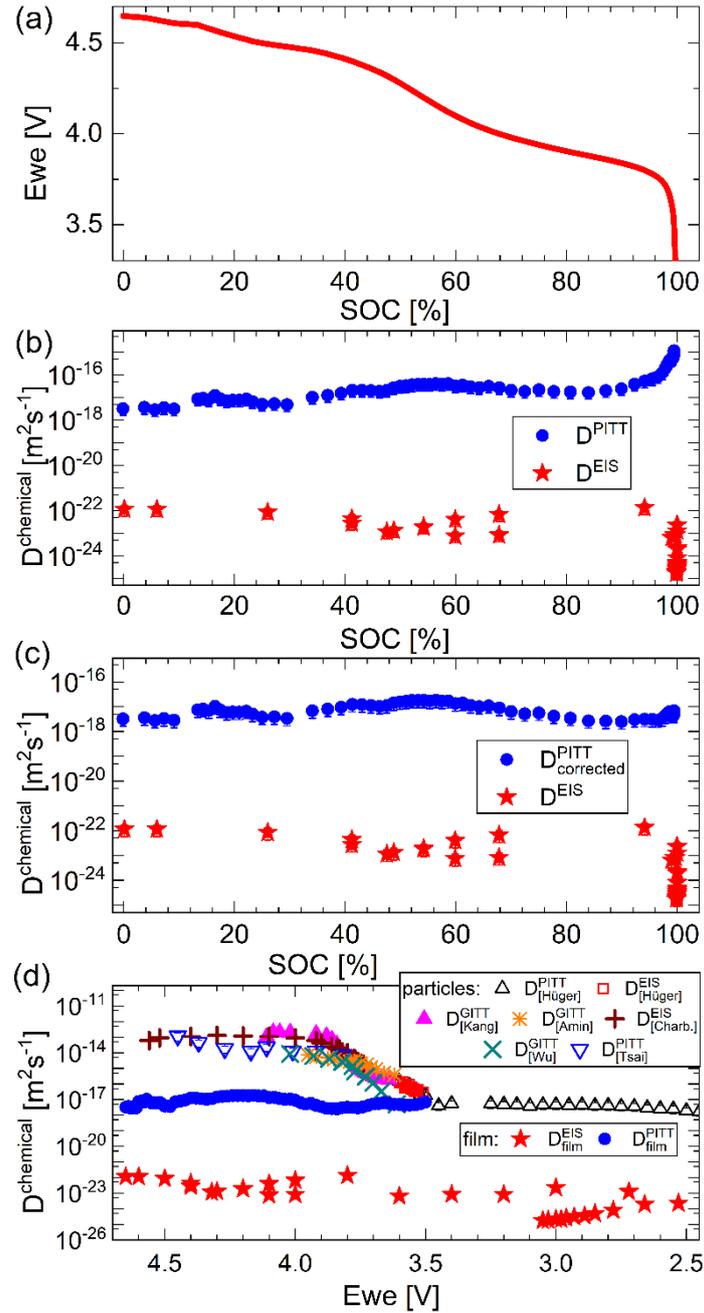

**Figure 7. (a)** Electrode potential vs. SOC obtained from PITT experiments performed during the first delithiation. **(b)** Li chemical diffusivities as obtained from EIS ($D^{EIS}$, marked with red stars) and from PITT experiments ($D^{PITT}$, marked with blue dots) as a function of SOC. **(c)** $D^{PITT}$ values corrected by taken in consideration equation (s17) and (s18) in the SM, compared to $D^{EIS}$ values, as a function of SOC. **(d)** $D^{EIS}$ and corrected $D^{PITT}$ obtained from the thin film compared to $D^{EIS}$ [19,22] and corrected $D^{PITT}$ [22,28], $D^{GITT}$ [29-31] values published for electrodes containing conglomerates of NMC111 particles or sintered pellets [22,30] as a function of electrode potential.

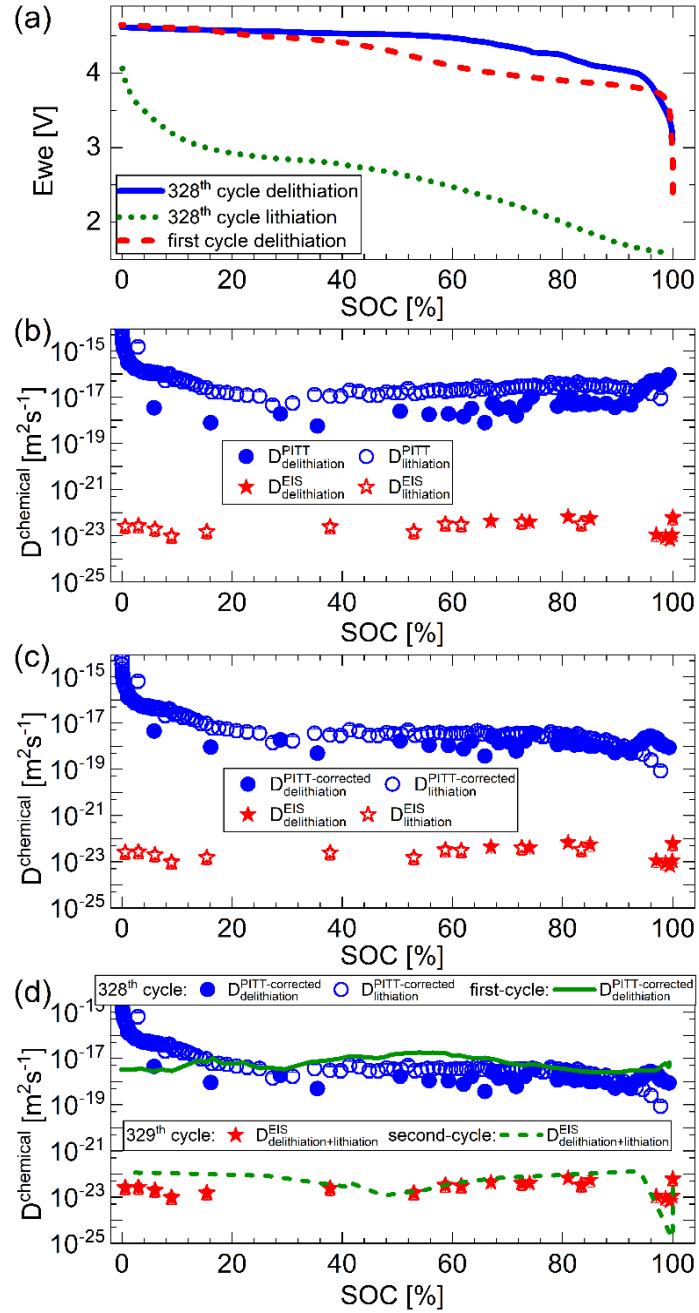

**Figure 8. (a)** Electrode potential vs. SOC obtained during PITT experiments performed during the 328[th] cycle (marked with blue line for delithiation and green dotted line for lithiation) compared to that of the first cycle (red dashed line, delithiation). **(b)** Chemical Li diffusivities obtained from EIS ($D^{EIS}$, marked with red stars) and from PITT experiments ($D^{PITT}$, marked with blue dots) as a function of SOC. **(c)** $D^{PITT}$ values corrected according to equation (s17) and (s18) (in the SM) compared to $D^{EIS}$ values, as a function of SOC. **(d)** $D^{EIS}$ and corrected $D^{PITT}$ values obtained from the 329[th] and 328[th] cycles, respectively, compared with the values obtained from the second cycle (for $D^{EIS}$, marked with dashed green line) and first cycle (for $D^{PITT}$, marked with solid green line).

The $D^{PITT}$ values are up to six orders of magnitude higher than the $D^{EIS}$ values. This discrepancy might be explained in the way these values were extracted from the measurement data. Details are provided in Section 7.4 of the SM. The $D^{PITT}$ values in Figure 7b and 8b were determined considering an electrode thickness of 1 µm. It is questionable whether the thickness used to determine $D^{PITT}$ should be the full thickness of the electrode. As recently shown, the EIS and PITT experiments performed on binder- and additive-free NMC111 bulk ceramics [22] could be brought into line only with help of a correction factor, in form of a restricted "diffusion depth" characteristic for a defined electrode. The corrected $D^{PITT}$ values were calculated in section 7.4 of the SM by introducing a reduced diffusion region due to the assumption of a specific diffusion mechanism. The results are shown in Figures 7c and 8c. Unfortunately, in the present case of films the modification is low and the discrepancy between $D^{PITT}$ and $D^{EIS}$ is still present. Note that the main message of the present paper is valid also without using the correction. Note also that the corrections based on the restricted "diffusion depth" has not been applied to the EIS results, since the determination of $D^{EIS}$ according to equations (s21,s22) in section 8 of the SM, is independent of the film thickness L.

The Li chemical diffusivities obtained in the present work are compared to literature values obtained from electrodes containing conglomerates of NMC111 particles in Figure 7d. These are (i) sintered NMC bulk samples without binder and conductive additives from GITT experiments performed by Amin et al. [30] and from EIS and PITT experiments performed by Hüger et al. [22], (ii) a single NMC111 particle from micro-PITT experiments performed by Tsai et al. [28], and (iii) a dried slurry thick film containing NMC111 particles together with binder and conductive additives from EIS experiments performed by Charbonneau et al. [19] and from the GITT experiments performed by Wu et al. [r29] and Kang et al. [31]. The correction of using equation (s8) was also applied to the literature $D^{PITT}$ and $D^{GITT}$ values shown in Figure 7d. The $D^{PITT}$ data of the present measurements (marked with blue dots in Figure 7d) are in agreement with the $D^{PITT}$ values of the sintered NMC111 pellet (black triangles in Figure 7d) at and before the onset of delithiation below 3.6 V, where we have virtually no Li extraction (see e.g. Figure 7a). When Li extraction starts (i.e., for potentials above 3.5 V, Figure 7a), the chemical diffusivities found in literature all increase strongly, while the $D^{PITT}$ values of the c-axis textured sputtered film electrodes remain at low values. This is consistent with the cycling results of Section 3.1, which showed much lower capacities for the NMC111 films than for electrodes made of agglomerated NMC111 particles.

The difference in chemical diffusivities during delithiation between literature data and the present results might be found in the structural and point defect state of the c-axis textured films. In classical untextured polycrystalline samples the structural state is modified in a way that diffusivities increase during delithiation. Possible reasons are a modification of Li vacancy concentration, of thermodynamic activity beyond the dilute limit or of the structure by Li extraction. This is not the case for the present films, where all quantities seem to be unaffected or single quantities compensate each other. Although, the reason of this difference is still unclear, the fact that chemical diffusivities are not increased during delithiation and keep close to the initial value is an important finding. Slow Li diffusion also during cycling is a crucial factor influencing capacities and cycling and may justify the behaviour of Figure 1.

Finally, Figure 8d shows the $D^{EIS}$ and corrected $D^{PITT}$ values obtained from the 329[th] and 328[th] cycles, respectively, in direct comparison with the $D^{EIS}$ and $D^{PITT}$ values obtained from the second and first cycles, respectively. There is good agreement between the PITT values on the one hand and the EIS values on the other. The diffusivities are almost independent of the SOC.

## 4.2 Discussion of Tracer Diffusivities

From EIS and PITT experiments the chemical diffusivity is determined, while from SIMS the Li tracer diffusivity. Chemical diffusivities describe ambipolar diffusion of Li$^+$ ions and electronic charge carriers [33]. In case of the electronic conductivity is orders of magnitude higher than the ionic conductivity, as in the present case [30,31], the Li tracer and chemical diffusivities are linked by the thermodynamic factor (TF) according to equation (s19) of the SM. The TF is calculated from the PITT data according to equation (s20). Its dependence on SOC and electrode potential is shown in Figure 9a for the first cycle and in Figure 10a also for the 328$^{th}$ cycle, respectively. The TF reaches very high values close to 100% SOC (at potentials below 3.5 V), because an extraction of a low amount of Li charge strongly increases the electrode potential (Figures 7a, 8a). Figures 9b and 10b show the Li tracer diffusivity obtained from PITT, EIS and SIMS studies as a function of SOC and electrode potential, respectively.

For the first cycle (Figure 9b) the tracer diffusivities determined from PITT and EIS decrease if the SOC is approaching 100% (Figure 9b). The $D^{PITT}$ values decrease by an order of magnitude from $D^{PITT} = 1 \times 10^{-20}$ m$^2$s$^{-1}$ at 99.0% SOC to $D^{PITT} = 1 \times 10^{-21}$ m$^2$s$^{-1}$ at 99.4% SOC. $D^{PITT}$ values at higher SOC (e.g. 100%) could not be measured because the method requires Li extraction. The decrease may continue above 99.4%, reaching values even below ≈ 1 × 10$^{-22}$ m$^2$s$^{-1}$ when SOC is close to 100%. This $D^{PITT}$ value is then in the range of the $D^{SIMS}$ values (the green squares in Figure 9b) corresponding to an SOC of 100%. Thus, the SIMS experiments confirm the $D^{PITT}$ values for the first cycle. In contrast, the $D^{EIS}$ values are significantly lower. This is also the case for the 328$^{th}$ cycle (Figure 10b). The $D^{SIMS}$ value obtained after cycling (marked with a filled green square in Figure 10b) agrees with the $D^{PITT}$ values, but by far not with the $D^{EIS}$ values. The true SOC position of $D^{SIMS}$ may not be exactly 100% as shown in Figure 10b because the re-lithiation process at the 329$^{th}$ cycle is most likely incomplete (see final voltage of 2.45 V in Figure S19). Therefore, the SOC corresponding to $D^{SIMS}$ may be slightly less than 100%. Only a slight deviation from 100% SOC to 98% SOC brings the $D^{SIMS}$ value in Figure 10b into perfect agreement with the $D^{PITT}$ values. Consequently, the higher Li tracer diffusivity of the cycled electrode compared to the pristine electrode is mainly due to an incomplete lithiation and a different corresponding SOC. In summary, SIMS suggests that the diffusivities obtained from PITT in the present work are more accurate than those obtained from EIS measurements. Note that the same extremely low $D^{EIS}$ are obtained when other diffusivity determination formulas are used (as shown in Section 8 of the SM, Figure S18a).

Regarding the EIS results, it should be mentioned that the tracer diffusivity of the pristine NMC film electrode, corresponding to the SOC = 100%, was calculated from the EIS measurements to $2.2 \times 10^{-29}$ m$^2$s$^{-1}$. This value is completely in agreement with the $D^{EIS}$ values obtained during the second cycle (Figure 9b). Details on the calculation of this value are provided in Section 8 of the SM.

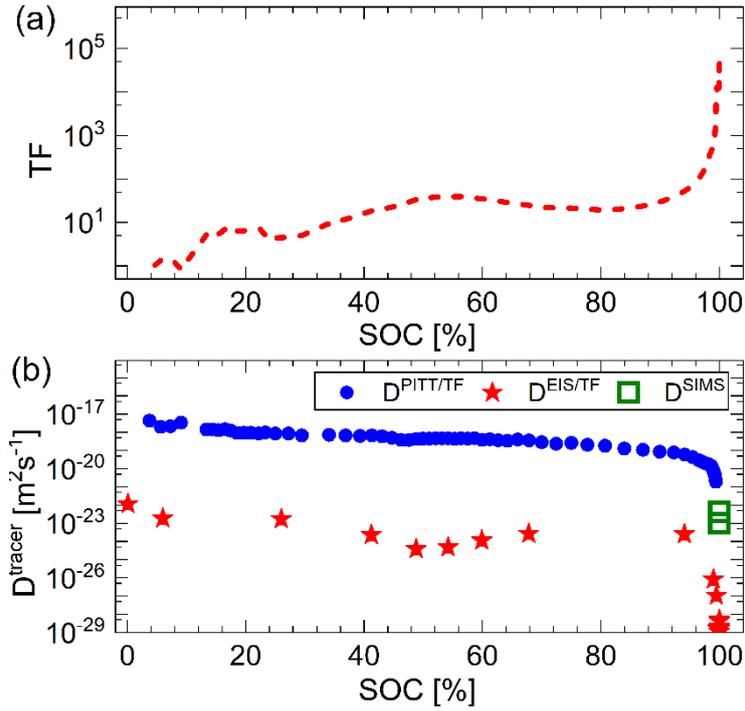

**Figure 9. (a)** Thermodynamic factor (TF) as obtained from the NMC111 film during the first cycle as a function of SOC. **(b)** Tracer Li diffusivities obtained from TF and chemical Li diffusivities $D^{EIS}$ and $D^{PITT}$ compared to those obtained from SIMS experiments as a function of SOC.

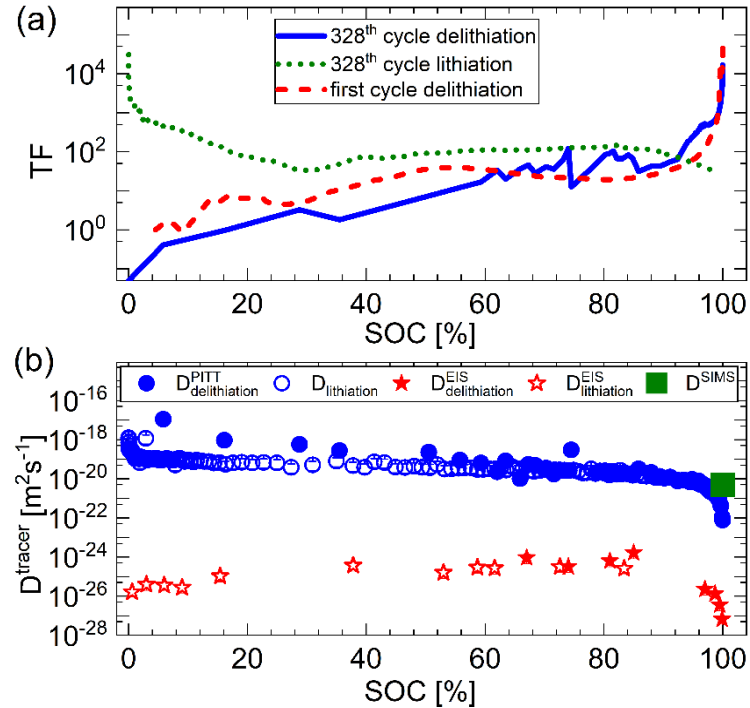

**Figure 10. (a)** Thermodynamic factor (TF) as obtained from the NMC111 film during the 328$^{th}$ cycle as a function of SOC. **(b)** Tracer Li diffusivities obtained from TF and chemical Li diffusivities $D^{EIS}$ and $D^{PITT}$ compared to those obtained from SIMS experiments after cycling as a function of SOC.

A main result of the present work is that the $D^{EIS}$ values (e.g. in Figures 9, 10) are significantly lower than the diffusivities determined from PITT and SIMS. A possible explanation is given in Figure 11. Figure 11 compares the $D^{EIS}$ and $D^{PITT}$ tracer diffusivities obtained in the current work on NMC111 films at room temperature with diffusivities measured on LCO single crystals where the diffusion was measured along the c-axis and perpendicular along a,b-direction in the pristine state [12]. Experimental results were extrapolated to room temperature. The result is that the $D^{PITT}$ value for the NMC111 film is consistent with the diffusivities in the ab-plane of the LCO single crystal, while the $D^{EIS}$ value is consistent with that of the c-axis oriented LCO single crystal. This indicates that by the three measurement methods, different diffusion process is probed. With EIS slow diffusion along the c-axis is measured while with PITT and SIMS faster diffusion in the a,b-plane is detected. For polycrystalline bulk NMC111 samples the tracer diffusivities detected by PITT, EIS and SIMS all agree and reach the value of the faster process (blue circle in Fig.11) [22], emphasizing the special role of thin NMC111 films.

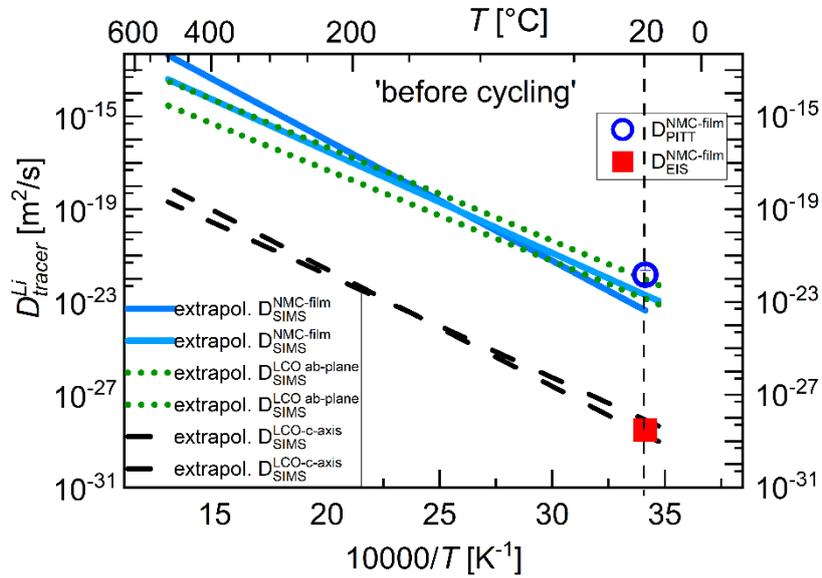

**Figure 11** Li tracer diffusivities calculated from the activation enthalpy and the pre-exponential factor for the c-axis textured NMC111 films [27] (blue solid lines), for LCO single crystals oriented along the c-axis (green dotted lines) and perpendicular to it (black dashed lines) [12]. The values at and near room temperature correspond to an extrapolation of the values measured at higher temperatures. The symbols indicate the Li tracer diffusivities at room temperature in the NMC films. They were obtained from the chemical diffusivities determined from EIS (marked with a filled square) and PITT (marked with an open circle) measurements using the thermodynamic factor determined from the PITT experiments.

As explained in Section 5 of the SM, the Li diffusion along c-axis is interpreted that even in the case where the Li enters the crystal along the c-axis, fast Li diffusion takes place predominantly within the ab-plane. The lower Li diffusivity along the c-axis occurs via rare defects in the transition metal oxide layer. These defects act as a kind of gate for Li diffusion to cross the transition metal oxide layer to reach the next lithium oxide layer, where fast diffusivity is again enabled. Therefore, the low diffusivities obtained in the c-axis direction are determined by the low density of defects in the transition metal oxide layer. Such defects might be defects in the transition metal layer. The reason for the different results of EIS and PITT is unclear at the moment.

### 4.3 Influence of Li Diffusivity on Electrode Specific Capacity

As shown in section 3.1, the specific capacity obtained during the C-rate capability experiment (Figure 1) and long-term cycling experiment (Figure 2) indicates that delithiation and lithiation do not take place within the whole film electrode of 1 µm. Only a limited part of the electrode at the interface to the electrolyte is relevant due to a low diffusivity. Depending on the current density used in the CC cycle experiment, the amount of charge density extracted (or inserted) during delithiation (or lithiation) is only a percentage of the maximum possible charge density of 66 µAhcm$^{-2}$ [18]. From this point of view, a mean Li diffusion length $d_i$ can roughly be calculated from each CC cycle

$$d_i = \frac{C_i}{C_{practical}} \cdot L. \tag{5}$$

Here, $C_i$ is the capacity at cycle $i$, $C_{practical}$ = 165 mAhg$^{-1}$, and $L$ = 1 µm is the film thickness. Figures 12a and 13a show the Li diffusion lengths calculated according to equation (5). The values obviously depend on the current density in such a way that cycling with lower current density increases the relevant diffusion length. This can be seen especially at the end of the cycling experiments (Figure 13a), i.e. by comparing cycles 326 and 327 with the previous ones (i.e. cycles 49 to 325). Cycles 49 to 325 were performed at a current density of 4 µAhcm$^{-2}$, whereas cycles 326 and 327 were performed at a current density of 0.4 µAhcm$^{-2}$, which is ten times lower. The estimated diffusion lengths (Figure 13a) for the lower current density is ten times longer than for the higher current density (Figure 13a). This occurs because the delithiation process takes place over a longer delithiation time interval at the lower current density. Consequently, for a given diffusivity, which is purely a material constant and independent of the current density $j_i$, the Li ions have more time $t_i$ to diffuse larger distances $d_i$ at lower currents. Using

$$D_i^{cycle} = \frac{d_i^2}{2 \cdot t_i} \tag{6}$$

the diffusivity ($D^{cycle}$) can be estimated. They are shown in Figures 12b and 13b with green squares as obtained from the C-rate capability experiment and from the long-term cycling experiment, respectively. From Figure 13b it can be observed that the diffusivity $D^{cycle}$ for cycle 325 (performed at a high current density of 4 µAhcm$^{-2}$) is identical to that of cycles 326 and 327 (performed at a low current density of 0.4 µAhcm$^{-2}$), which supports further the idea that the specific capacities determined in the present work on NMC111 films (Figures 1,2) are determined by diffusion. Although the current density is very different, the same diffusivity is obtained. The $D^{cycle}$ values for the other cycles with different current densities (Figures 12b,13b) are all the same within one order of magnitude, indicating approximately the same diffusivity within error limits. Therefore, $D^{cycle}$ appears to be independent of current density and independent of the number of cycles. This is a similar behavior as discussed for diffusivities obtained from PITT, EIS and SIMS.

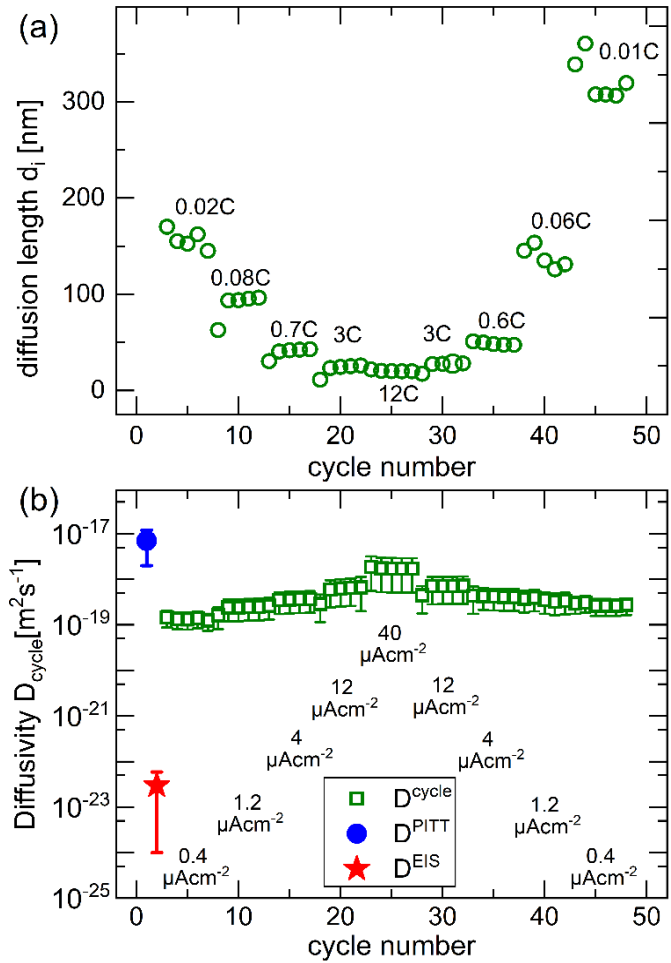

**Figure 12.** (a) Diffusion lengths (marked with green open circles) and (b) diffusivities (marked with green open squares, $D^{cycle}$) determined from the discharge capacities of the C-rate capability experiment (see Figure 1) using equations (5) and (6), respectively. The blue point and the red star represent the mean value of the diffusivities obtained from the PITT experiment during the first cycle and from the EIS experiments during the second cycle, respectively.

Note that the $D^{cycle}$ values are rough estimates averaged over the whole cycle. This means that in order to compare $D^{cycle}$ with $D^{PITT}$ and $D^{EIS}$, one has to calculate also the averages of the latter two quantities over the whole cycled NMC film electrode potentials. This can be done correctly because $D^{PITT}$ and $D^{EIS}$ show only a modest dependence on SOC and Ewe during cycling (Figures 7c,d, 8c,d, 9b, and 10b). As a result, we use the average $D^{PITT}$ and $D^{EIS}$ chemical diffusivities for further discussion. The blue dots in Figures 12b, 13b show the mean diffusivities obtained by PITT during the first and $328^{th}$ cycle. The red stars in Figures 12b,13b show the mean values of the diffusivities obtained by EIS during the second and the $329^{th}$ cycle. It can be observed that the $D^{cycle}$ values are much closer to the diffusivities obtained by PITT and significantly higher than those obtained by EIS. Consequently, the $D^{cycle}$ values confirm the results of the SIMS experiments that the $D^{PITT}$ values are more reliable than the $D^{EIS}$ values.

This reinforces the idea that the low $D^{EIS}$ values reflect only slow diffusion processes, whereas the $D^{PITT}$, $D^{SIMS}$ and $D^{cycle}$ values capture the faster diffusion relevant for electrochemical cycling.

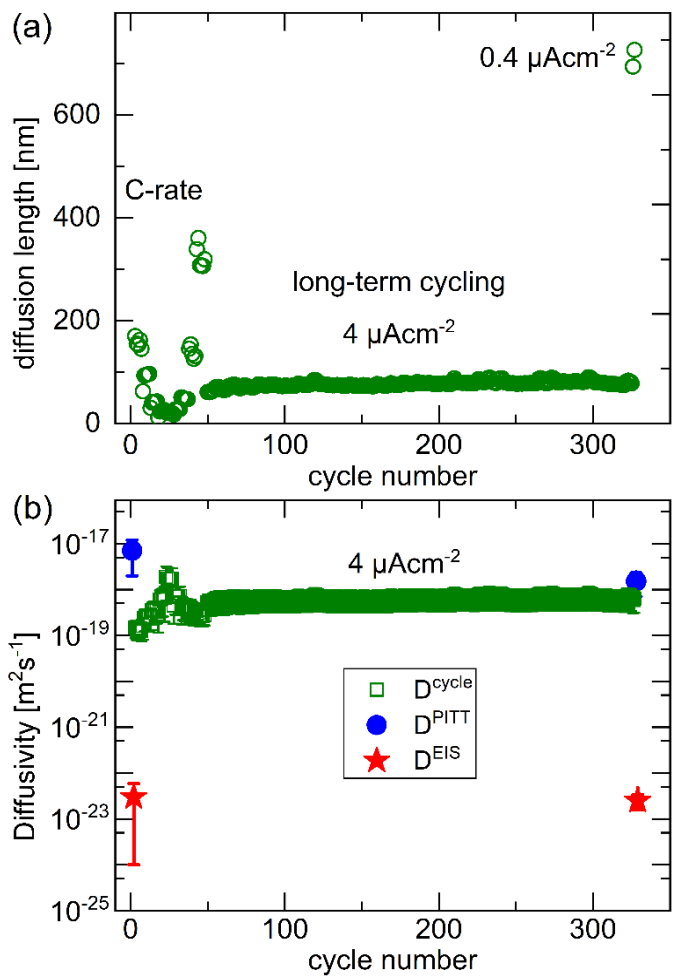

**Figure 13.** (a) Diffusion lengths (marked with green open circles) and (b) diffusivities (marked with green open squares, $D^{cycle}$) determined from the discharge capacities of the C-rate capability experiment (up to cycle 48, see Figure 1) and the long-term cycling experiment (starting from cycle 49, see Figure 2) using equations (5) and (6), respectively, The blue point and the red star represent the mean value of the diffusivities obtained from the PITT experiment during the first cycle and the 328$^{th}$ cycle, and from the EIS experiments during the second cycle and the 329$^{th}$ cycle, respectively. Note that the last two cycles of the long-term cycling experiment, i.e. cycles 326 and 327, were done with a current density of 0.4 µAhcm$^{-2}$, which is ten times lower than that of the other cycles between cycle 49 and cycle 325.

Another aspect concerning the influence of the Li diffusivity on the gravimetric capacities of NMC111 films is revealed by using the measured Li chemical diffusivities of Figure 8d to calculate the highest C-rate at which Li diffusion over the full depth of the NMC111 film is possible. This will allow full delithiation and lithiation, and the gravimetric capacity reaches its maximum value. The further analysis of the chemical diffusivities is limited to the $D^{PITT}$ values of the 328$^{th}$ cycle. The values determined from delithiation are slightly lower than those from lithiation (Figure 13c,d), and it is expected that they will determine the C-rate. Therefore, we consider the mean $D^{PITT}$ chemical diffusivity of the 328$^{th}$ cycle during delithiation, which is 1.5 × 10$^{-18}$ m$^2$s$^{-1}$. According to t = L$^2$/(2D$^{PITT}$), where L = 1 µm, a mean time interval of t ≈ 93 hours and thus a C-rate of (1/93) ≈ 0.01C is obtained. If we also consider a $D^{PITT}$ error of about 50%, a lower diffusivity of 7.5 × 10$^{-19}$ m$^2$s$^{-1}$ can be taken into account. In this case, a C-rate of ≈ 0.005C is obtained. Thus, the upper limit of the C-rate for which Li diffusion can take place in the entire depth of the film electrode is between 0.01C and 0.005C.

This result explains why the gravimetric capacity measured during the C-rate experiment (Figure 1) and that during the long-term experiment (Figure 2) reach only a percentage of the expected capacity for the NMC, even for small C-rates down to ≈ 0.01C. Based on the above considerations, the C-rate has to be below 0.01C to allow complete Li transport over the full depth of the film electrode. The delithiation process during the PITT experiment of the first cycle and of the 328$^{th}$ cycle was 113 h and 187 h, respectively, corresponding to a sufficiently low C rate of ≈ 0.0083 C and ≈ 0.0053 C, allowing, as also measured, almost complete delithiation over the entire film thickness.

## 5. Conclusions

In summary, the following results were obtained from the experiments performed in this study:

1. The present work provides evidence for the widely held assumption that the diffusion of Li in electrode materials of LIBs can be the determining factor governing the insertion and removal of Li into/from electrodes during cycling. In other words, the Li diffusivity is a crucial kinetic parameter that determines how fast Li can be removed and reinserted into the electrode. Consequently, the diffusivity determines the achieved specific capacity at certain current densities. This was exemplified on c-axis textured NMC111 thin film electrodes, where thin and flat films provide a convenient platform for studying the properties of active materials used in LIBs. Evidence for these claims was provided by (i) measuring the behavior of the capacity in C-rate capability and long-term cycling experiments versus current densities, (ii) estimating the diffusivity from these capacities, and (iii) comparing the estimated diffusivities with diffusivities obtained from PITT, EIS, and SIMS experiments.

2. The C-rate capability and long-term cycling behavior of up to 1 µm thick NMC111 films deposited by ion beam sputtering on polished stainless steel current collectors were investigated. Polycrystalline films were obtained by post-annealing the films for 1 h at 700 °C. It was observed that these electrodes exhibited a cycling behavior with reduced specific capacity. The capacity showed a reversible decrease with increasing current density, indicating no film degradation. This phenomenon was attributed to slow Li diffusion, which significantly limits the specific capacity at a given current density.

3. During cycling, the films reached only ≈ 30% of the expected gravimetric capacity (165 mAhg$^{-1}$), even at low C-rates such as 0.01C. Slower cycling with C-rates below 0.01C, as in long-term PITT experiments, allowed high capacities for the delithiation (charging) process, even above 200 mAhg$^{-1}$. The capacity decreased reversibly with increasing current density, again indicating no film destruction during delithiation and lithiation cycles between 1.6 V and 4.6 V, even after 329 cycles.

4. The Li kinetics was studied by determining the Li chemical diffusivities using the electrochemical-based measurement techniques of EIS and PITT. These diffusivities were compared to Li tracer diffusivities using the SIMS technique. The chemical and tracer diffusivities were linked by the thermodynamic factor (TF) from the PITT experiments. SIMS confirmed that the diffusivities obtained from PITT are more relevant for electrochemical cycling than the lower values obtained from EIS measurements.

5. Before the onset of significant delithiation during the first cycle, the chemical diffusivities obtained by PITT on NMC111 films are similar to those obtained on sintered NMC111 bulk electrodes. During cycling, the chemical diffusivities of the films are not significantly dependent on the SOC. Low diffusivities are present during cycling, making Li diffusion the dominating process governing Li incorporation and release in NMC111 films.

6. SIMS experiments were also performed to determine the temperature dependence of the Li tracer diffusivities in cycled NMC111 film electrodes. The diffusivities of the cycled NMC111 film electrode are higher than those of a pristine electrode, mainly due. to the lower activation enthalpy of diffusion of $(0.65 \pm 0.02)$ compared to $(0.95 \pm 0.08)$ eV.

7. The discrepancy between the $D^{EIS}$ values on the one hand and $D^{PITT}$ and $D^{SIMS}$ on the other hand can be explained by the assumption that different processes are probed. EIS probes slow diffusion along the c-axis, while PITT and SIMS measure faster diffusion in the a,b-plane. The latter process governs electrochemical cycling.

8. The low Li diffusivities determined are responsible for the low measured gravimetric capacity. The C-rate experiments show that the increase in C-rate is proportional to the square root of the increase in current density, which, indicates that slow Li diffusion is responsible for the capacity decrease at higher current densities. The similar charge density and the doubled gravimetric capacity in 500 nm thin films compared to 1 μm thin films supports this. Based on the percentage of specific capacitance achieved during the C-rate and long-term cycling experiments, diffusivities were estimated from the cycling experiments. The estimated diffusivities were roughly independent of the current density and in agreement with the diffusivities measured by PITT and SIMS indicating again that the specific capacity is determined by the diffusivity. A C-rate limit for full delithiation below 0.01 C was calculated, which agrees with that obtained from cycling experiments.

**Acknowledgements**

This work was funded by the Deutsche Forschungsgemeinschaft (DFG, German Research Foundation) under the contract SCHM 1569/33-2 (413672097). The financial support is gratefully acknowledged. We are indebted to Gerrit Zander for ICP-OES analysis and Reinhard Deichmann for producing the sputter targets.

AUTHOR DECLARATIONS

**Conflict of Interest**

The authors have no conflicts to disclose.

**Author Contributions**

**Erwin Hüger**: Investigation (lead); Formal analysis (lead); Validation (lead); Visualization (lead); Writing – original draft (lead); Conceptualization (equal); Data curation (equal); Methodology (equal); Writing – review & editing (equal). **Harald Schmidt**: Funding Acquisition (lead); Project Administration (lead); Conceptualization (equal); Data curation (equal); Methodology (equal); Formal analysis (supporting); Validation (supporting); Visualization (supporting); Writing – review & editing (equal).

DATA AVAILABILITY

The data that support the findings of this study are available from the corresponding author upon reasonable request.